\documentclass[10pt,journal]{IEEEtran}
\usepackage{nomencl}
\usepackage{graphicx}
\usepackage{amsmath,amssymb,amsthm, amstext}
\usepackage{epsf}
\usepackage[T1]{fontenc}
\usepackage{subfigure}
\usepackage{algorithm}
\usepackage{algorithmic}
\usepackage{url}
\usepackage{enumerate}

\newtheorem{definition}{Definition}
\newtheorem{lemma}{Lemma}

\newtheorem{remark}{Remark}
\newtheorem{proposition}{Proposition}

\newcommand{\be}{\begin{eqnarray}}
\newcommand{\ee}{\end{eqnarray}}
\newcommand{\bearno}{\begin{eqnarray*}}
\newcommand{\enarno}{\end{eqnarray*}}
\newcommand{\dref}[1]{(\ref{#1})}

\begin{document}
\title{Towards a Theory of Anonymous Networking}
\author{\IEEEauthorblockN{J. Ghaderi and R. Srikant\\}
\IEEEauthorblockA{Department of ECE, and Coordinated Science Lab.\\
University of Illinois at Urbana-Champaign\\
\{jghaderi, rsrikant\}@illinois.edu}
\thanks{This research was supported by NSF Grant 07-21286 and an ARMY MURI subcontract.}
\thanks{An earlier version of the paper has been appeared in the Proceedings of IEEE INFOCOM 2010.}}
\maketitle
\begin{abstract}
The problem of anonymous networking when an eavesdropper observes packet timings in a communication network is considered. The goal is to hide the identities of source-destination nodes, and paths of information flow in the network. One way to achieve such an anonymity is to use \textit{Mixes}. Mixes are nodes that receive packets from multiple sources and change the timing of packets, by mixing packets at the output links, to prevent the eavesdropper from finding sources of outgoing packets. In this paper, we consider two simple but fundamental scenarios: double input-single output Mix and double input-double output Mix. For the first case, we use the information-theoretic definition of the anonymity, based on average entropy per packet, and find an optimal mixing strategy under a strict latency constraint. For the second case, perfect anonymity is considered, and maximal throughput strategies with perfect anonymity are found under a strict latency constraint and an average queue length constraint.
\end{abstract}
\section{Introduction}
Secure communication has become increasingly important. Privacy and anonymity considerations apply to all components of a communication network, such as contents of data packets, identities of source-destination nodes, and paths of information flow in the network. While a data packet's content can be protected by encrypting the payload of the packet, an eavesdropper can still detect the addresses of the source and the destination by \textit{traffic analysis}. For example, observing the header of the packet can still reveal the identities of its corresponding source-destination pair. \textit{Onion Routing} \cite{onion} and \textit{Tor} network \cite{tor} are well-known solutions that provide protection against both eavesdropping and traffic analysis. The basic idea is to form an overlay network of Tor nodes, and relay packets through several Tor nodes instead of taking the direct path between the source and the destination. To create a private network, links between Tor nodes are encrypted such that each Tor node only knows the node from which it receives a packet and the node to which it forwards the packet. Therefore, any node in the Tor network sees only two hops (the previous and next nodes) but is not aware of the whole path between the source and the destination, Therefore, a compromised node cannot easily identify source-destination pairs. But Tor cannot solve all anonymity problems. If an eavesdropper can observe the traffic in and out of some nodes, it can still correlate the incoming and outgoing packets of relay nodes to identify the source and the destination or, at least, discover parts of the route between the source and the destination. This kind of statistical analysis is known as timing analysis since the eavesdropper only needs packet timings. For example, in Figure \ref{mixer}, if the processing delay is small, there is a high correlation between output and input processes, and the eavesdropper can easily identify the source of each outgoing packet.

 To provide protection against the timing analysis attack, nodes in an anonymous network need to perform an additional task, known as mixing, before transmitting packets on output links. A node with mixing ability is called a \textit{Mix}. In this solution, a Mix receives packets from multiple links, re-encrypts them, and changes the timings of packets, by mixing (reordering) packets at the output links, in such a way that the eavesdropper cannot relate an outgoing packet to its corresponding sender.

  The original concept of Mix was introduced by Chaum \cite{chaum}. The Mix anonymity was improved by random delaying \cite{kes} (Stop-and-go MIXes), and dummy packet transmission \cite{pfitz} (ISDN-MIXes), and used for the various Internet applications such as email \cite{gulcu} and WWW \cite{reiter}(Crowds). Other proposed anonymity schemes are JAP \cite{jap}, MorphMix \cite{morph}, Mixmaster \cite{master}, Mixminion \cite{minion}, Buses \cite{buses}, etc.

   However, theoretical analysis of the performance of Chaum mixing is very limited. The information-theoretic measure of anonymity, based on Shannon's equivocation \cite{shannon}, was used in \cite{diaz}, \cite{andrei} to evaluate the performance of a few mixing strategies, under some attack
scenarios, however, their approach does not take into account the delay or traffic statistics; whereas, modifying packet timings to obfuscate the eavesdropper indeed increases the transmission latency.
In \cite{diaz2}, \cite{diaz3}, the authors consider the performance
of so-called generalized Mixes, e.g., a timed pool Mix, under some active attack scenarios.
In a timed pool Mix, the Mix collects input messages that are placed in a pool during a
round/cycle of the Mix, and then flushes the messages out with some
probability. The flushing probability is characterized by a function
$P(n)$ representing the probability of the messages being sent in the
current round, given that the Mix contains $n$ messages in the pool.
Again, this is a very specific type of Mix and it is not clear if this
is optimal. So, the question of interest is the following: what is the maximum achievable anonymity under a constraint on delay?

  Characterizing the anonymity as a function of traffic load and the delay constraint has been considered in \cite{prav}. The authors in \cite{prav} have considered a Mix with two input links and one output link, where arrivals on the input links are two poisson processes with equal rates, and they characterize upper and lower bounds on the maximum achievable anonymity under a strict delay constraint. The basic idea is that the Mix waits for a certain amount of time, collects packets from two sources, and sends a batch containing the received packets to the output. The implicit assumption in \cite{prav} is that there is no constraint on the capacity of the output link, i.e., the batch can be transmitted instantaneously at the output, no matter how many packets are contained in the batch.

\begin{figure}[!t]
\centering
\includegraphics [width = 2.4 in]{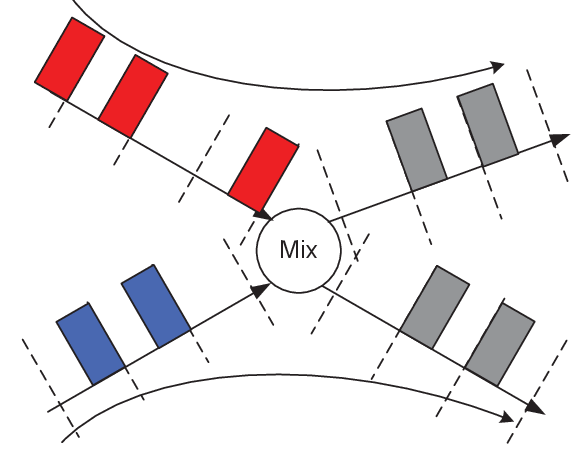}
\caption{The double input-double output Mix. The capacity of each link is $1$ packet/time slot.}
\label{mixer2}
\end{figure}

  The path between any source-destination pair in an anonymous network contains several nodes; each of which has, possibly, several input links and several output links. At each node, to perform routing, traffic generated by two or more sources can be merged into one outgoing stream, or the merged stream can be decomposed at several output links for different destinations. To expose the main features of mixing strategies, we focus on two fundamental cases: a double input-single output Mix, Figure \ref{mixer}, and a double input-double output Mix, Figure \ref{mixer2}. Compared to \cite{prav}, our model considers cases with finite link capacities and we derive optimal solutions for certain cases. The remainder of the paper is organized as follows. In section \ref{double-single}, the double input-single output Mix is considered, and the optimal mixing strategy is found to maximize the anonymity under a strict latency constraint. Section \ref{double-double} is devoted to the double input-double output Mix, where the optimal mixing strategy is found under a strict latency constraint and an average queue length constraint. Finally, we end the paper with some concluding remarks.

\section{Double Input-Single Output Mix}\label{double-single}
\begin{figure}[!t]
\centering
\includegraphics [width = 2.5 in]{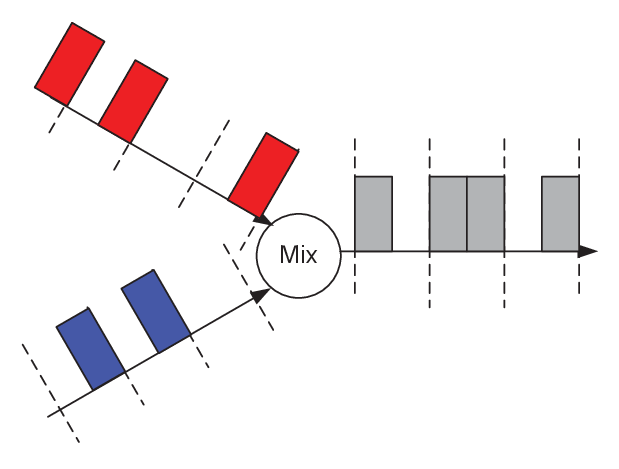}
\caption{The double input-single output Mix. The capacity of each input link is $1$ packet/time slot and the capacity of the output link is $2$ packets/time slot.}
\label{mixer}
\end{figure}
Consider Figure \ref{mixer} where there are two incoming flows, red and blue, and one outgoing link. The capacity of each input link is $1$ packet/time-slot, and the capacity of the output link is $2$ packets/time-slot. This model ensures that packets do not have to be dropped due to lack of capacity, even when the input links bring in data at maximum rate. Red and blue packets arrive according to i.i.d. Bernoulli processes with rates $\lambda_R$ and $\lambda_B$ respectively. There is an eavesdropper observing the incoming and outgoing packets. Assume the eavesdropper knows the source of each incoming packet, i.e., its color. This might be made feasible by traffic analysis if the Mix is the first hop of the route or, otherwise, by timing analysis of the previous hop. Given the source of each incoming packet, the eavesdropper aims to identify the source of each outgoing packet, i.e., assign colors, red and blue, to the outgoing stream of packets.

First, consider the case where we do not allow for any delay, i.e., the Mix must send packets out in the same slot in which they arrived. Note that this is possible, without any packet drop, since at most two packets arrive in each slot, and the capacity of the output link is $2$ packets/slot. Then, the only way to confuse the eavesdropper is to send out a random permutation of received packets in each slot.

By allowing a strict delay $T \geq 1$ for each packet, the Mix can do better; it can select and permute packets from the current slot and also from the previous slots, up to $T$ slots before.

Next, we introduce a few notations and then define the mixing strategy under a strict delay constraint $T$ per packet. The definitions are quite general and could be used to describe the mixing strategy and its anonymity for any multiple input-single output Mix, under a delay constraint $T$.

Let the random variable $I_k$ denote arrivals in $k$-th slot. Therefore, for the double input-single output Mix, $I_k$ can be $\emptyset$, $R$, $B$, or $RB$, where they respectively denote the cases of no arrivals, red arrival but no blue arrival, blue arrival but no red arrival, and both red and blue arrivals.
Similarly define a random variable $O_k$ for the output at slot $k$. For the double input-single output Mix, $O_k \in \left\{\emptyset, R, B, RR, BB, RB, BR \right\}$ (note that ordering of $RB$ or $BR$ at the output matters). There is an eavesdropper (Eve) who knows whether there is a packet at the output or not, i.e., at each time $k$, Eve knows a random variable $G_k := |O_k|$. For any stochastic process $\{X_k\}_{k \geq 1}$, define $X^{(N)} :=  X_1 \cdots X_N$.

Since we work with a discrete-time system, we have to specify the order in which different events occur: at each time slot $k$, first packet arrivals occur at the beginning of the time slot, and then departures occur.
At the end of time slot $k$, let $Q_{k}(j)$ be a queue containing packets that have been in the system for $j$ time slots, $0 \leq j \leq T-1$ ($j=0$ corresponds to packets arrived in the current slot but have not transmitted). In the case of double input-single output Mix, obviously, $Q_{k}(j) \in \{\emptyset, R, B, RB\}$, for $0 \leq j \leq T-1$. Let $Q_{k} := [Q_{k}(0), Q_{k}(1), \cdots Q_{k}(T-1)]$ be the collection of such queues which represent existing packets in the Mix at the end of slot $k$. A mixing strategy consists of a \textit{selection} strategy followed by a \textit{permutation} strategy. At each time $k$, the Mix randomly selects packets from $Q_{k-1}$, and also from the newly arrived packets $I_k$, and then send a random permutation of the selected packets to the output. Let $D_k(j+1)$ denote the packets selected from $Q_{k-1}(j)$ by the Mix at slot $k$, for $0 \leq  j \leq T-1$\footnote{At the moment that we move from slot $k-1$ to slot $k$, the delay of each remaining packet increases by $1$, hence we have used the notation $D_k(j+1)$ for packets selected from queue $Q_{k-1}(j)$ at time $k$.}. Then, the queue dynamics can be described as
\begin{equation}
Q_{k}(j+1) = Q_{k-1}(j) \backslash D_{k}(j+1),
\end{equation}
and
\begin{equation}
Q_k(0)= I_k \backslash D_k(0)
\end{equation}
where $D_k(0)$ denoted the packets selected from new arrivals $I_k$ and ``$\backslash$'' is the set difference operator. Note that $Q_{k}(T)=\emptyset$  because for any mixing strategy, under delay constraint $T$, packets that have been in the system for $T$ time slots have to be transmitted, i.e., $D_k(T)=Q_{k-1}(T-1)$. Let $D_k:=[D_k(0), \cdots D_k(T)]$. Hence, $D_k=\Upsilon(Q_{k-1},I_k)$, for some random selection strategy $\Upsilon$, and $O_k=\Pi(D_k)$ for some permutation strategy $\Pi$, thus $O_k=\psi(Q_{k-1},I_k)$ where $\psi= \Pi \circ \Upsilon$ is the mixing strategy. Let $\Psi_T$ denote the set of all possible mixing strategies that satisfy the strict delay constraint $T$.

 Next, we define anonymity of a mixing strategy $\psi \in \Psi_T$, based on the average conditional entropy of the output sequence given the input sequence and the sequence $G$, as follows.
\begin{definition}
The anonymity $A^\psi$ of a mixing strategy $\psi$ is defined as
\begin{equation}\label{def}
A^\psi= \lim_{N \to \infty}\frac{1}{N(\lambda_R+\lambda_B)}H(O^{(N)}|I^{(N)},G^{(N)}).
\end{equation}
\end{definition}
Note that in the above definition, the numerator is the entropy of the output sequence of length $N$ given that Eve: (i) observes the input sequence of length $N$, and (ii) detects packet transmissions at the output. The denominator is the average number of red and blue arrivals in $N$ slots. So, as $N \to \infty$, anonymity is the amount of uncertainty in each outgoing packet, bits/packet, observed by Eve.
\begin{remark}
By using the Fano's inequality, the anonymity provides a lower bound for the probability of error in detection incurred by the eavesdropper \cite{prav2}.
\end{remark}
Without loss of generality, we can assume that the optimal mixing strategy does not change the ordering of packets from the same flow. This can be justified as follows. For any input sequence of length $N$, consider the set of possible output sequences of length $N$, under the mixing strategy $\psi$. If, for example, the red packets do not appear at the output in the same order as their arrival order, we can simply reorder them according to their arrival order, without changing the conditional probability of the appearance of the output sequence given the input sequence. Note that this does not reduce the anonymity and also will not violate the delay constraint (Eve is only interested in detecting the colors of the packets at the output, not their arrival times.). This also means that the mixing strategy is compatible with network protocols such as TCP, as it does change the sequence numbers of packets from the same flow.
\subsection{Structure of Optimal Mix Strategy}\label{structure}
We wish to find the optimal strategy $\psi^* \in \Psi_T$ that maximizes the anonymity.
For mixing under a strict delay $T$, the output packets at time $k$ are selected from the packets received in the current slot or the past $T$ time slots that have not been transmitted before; however, in general, the rule, which determines which packet has to be transmitted, itself could possibly depend on the entire history up to time $k$. In other words, if we let $\mathcal{F}_k := \{I^{(k-1)},\ O^{(k-1)}\}$ denote the history up to the beginning of time slot $k$, $O_k=\psi_{k,\mathcal{F}_k}\big(Q_k,I_k)$.
Here, we assume that the Mix strategy and the Eve detection strategy are both causal and online meaning that transmission and detection decisions for $O_k$, has to be made at time $k$, and only depend on the history up to time $k$. Then, we show that the optimal mixing strategy, in fact, only needs a finite memory $T$ and does not need the entire history, i.e., $O_k=\psi _{Q_{k-1}}\big(Q_{k-1},I_k\big)$.

By the chain rule \cite{cover}, the conditional entropy in (\ref{def}) can be written as
\begin{align}
& H(O^{(N)}|I^{(N)},G^{(N)}) \nonumber \\
& = \sum_{k=1}^NH(O_k|I^{(N)},G^{(N)},O^{(k-1)}) \label{true}\\
& = \sum_{k=1}^N H(O_k|I^{(k)},G^{(k)},O^{(k-1)}) \label{causality} \\
& = \sum_{k=1}^N H(O_k|I^{(k)},G_k,O^{(k-1)}) \label{Gdef}\\
& = \sum_{k=1}^N H(O_k|I^{(k)},G_k,O^{(k-1)},Q_{k-1}) \label{upper}\\
& \leq \sum_{k=1}^N H(O_k|G_k, Q_{k-1},I_k), \label{upper-Q}
\end{align}
where (\ref{causality}) is because the Mix strategy and the Eve detection strategy are both causal and online, (\ref{Gdef}) follows from the fact that given $O^{(k-1)}$, $G^{(k-1)}$ does not contain any information, \dref{upper} holds because given $I^{(k-1)}$ and $O^{(k-1)}$, $Q_{k-1}$ is known, and finally \dref{upper-Q} follows from the fact that conditioning reduces the entropy.

Hence, for any mixing strategy $\psi \in \Psi_T$,
\begin{align} \label{upper2}
A^{\psi} \leq  \lim_{N \to \infty}\frac{1}{N(\lambda_R+\lambda)}\sum_{k=1}^N H(O_k|G_k, Q_{k-1},I_k).
\end{align}
Next, consider maximizing the Right-Hand-Side (RHS) of (\ref{upper2}).
We can define $Q_{k}$ to be state of a Markov chain $Q$ at the end of time $k$ (at the beginning of time $k+1$). Note that, for the double input-single output Mix, under strict delay $T$, $Q_{k}\in  \{\emptyset, R,B,RB\}^{T}$. Hence, maximizing RHS of (\ref{upper2}) can be interpreted as an average reward maximization problem over an appropriately defined \textit{Markov Decision Process} (MDP) $Q_k$ with finite number of states. More accurately, maximizing RHS of (\ref{upper2}) can be written as
\be \label{upper3}
\max_{\psi \in \Psi_T} \lim_{N \to \infty}\frac{1}{N}\sum_{k=1}^N \mathbb{E}_{q_{k-1}}[{c(q_{k-1},\psi_k)}],
\ee
where the reward function is given by
\be
c(q_{k-1},\psi_k) & = & H(O_k|q_{k-1}, I_k, G_k)\nonumber \\
&=&\mathbb{E}_{i_k}[{H(O_k|q_{k-1}, i_k, G_k)}],
\ee
where $q_k$ and $i_k$ are realizations of random variables $Q_k$ and $I_k$ respectively\footnote{Throughout the paper, we use a capital letter for a random variable and the corresponding lower-case letter to denote a realization of that random variable.}. Hence, we have a finite state MDP with bounded rewards. Consider the class of stationary (and Markov) policies $S_T \subset \Psi_T $, where, at each time $k$, decision is made only based on $Q_{k-1}$, i.e., $O_k=\psi _{Q_{k-1}}\big(Q_{k-1},I_k\big)$. A sufficient condition for existence of an optimal stationary policy for \dref{upper3} is given by the following lemma~\cite{bert,kumar}.
\begin{lemma}\label{mdp}
Suppose there exists a constant $w$ and a bounded function $\phi$, unique up to an additive constant, satisfying the following optimality equation
\be\label{dp}
w+\phi(q)=\max_u \left\{c(q,u)+E[\phi(Q_1)|Q_0=q, \psi_1=u]\right\},
\ee
then $w$ is the maximal average-reward and the optimal stationary policy $\psi^*$ is the one
that chooses the optimizing $u(.)$.
\end{lemma}
Next, we show that \dref{dp} always has a solution. Let $\tau := \min \{k \geq 1: Q_k=\emptyset\}$ where by $Q_k=\emptyset\equiv \{\emptyset\}^T$ we mean a state where there are no packets in the Mix waiting for transmission. Next, consider the class of stationary policies $S_T \subset \Psi_T $, i.e., $O_k=\psi _{Q_{k-1}}\big(Q_{k-1},I_k\big)$. Then we have, for any $\psi \in S_T$, and for any initial state $q$,
\be \label{return}
\mathbb{E}^\psi[\tau|Q_0=q] \leq \frac{T}{(1-\lambda_R)^{T}(1-\lambda_B)^{T}} < \infty.
\ee
This is because starting from any initial state, after a null input sequence of length $T$, i.e., no arrivals for $T$ time slots, the Markov chain has to return to $\emptyset$ state. Such a sequence occurs with probability $(1-\lambda_R)^{T}(1-\lambda_B)^{T}$ and the expected time is bounded as in \dref{return} for any stationary Markov policy as long as $\lambda_R$ and $\lambda_B$ are strictly less than one. The following lemma is a corollary of Theorem (6.5) of \cite{kumar} (page 164).
\begin{lemma}\label{existance}
Assume $\mathbb{E}^\psi[\tau|Q_0=q]\leq B < \infty$ for all $\psi \in S_T$ and all $q$, then there exists a $w$ and $\phi(.)$ satisfying DP equation \dref{dp}.
\end{lemma}
Hence, it follows from Lemmas \ref{existance} and \ref{mdp} that the maximizer of RHS of (\ref{upper2}) is a stationary policy. Moreover, observe that for any stationary policy, the inequality in \dref{upper-Q} and accordingly in \dref{upper2} can be replaced by equality, and therefore, the stationary policy $\psi^*$ given by \dref{dp} always exists and actually maximizes the anonymity and the maximum achievable anonymity, by definition, is given by
$$
A^{\psi^*}=\frac{w}{\lambda_R+\lambda_B}.
$$
Note that the above argument is still valid for any multiple input-single output Mix and any link capacities as long as the output link capacity is greater than or equal to the sum of input-links' capacities to ensure that there is no need to drop packets.

In general, the DP equation \dref{dp} can be written in form of the following LP
\begin{align}
&\max \ w & \\
&w+\phi(q) \leq c(q,u)+E[\phi(Q_1)|Q_0=q, u_1=u]\ ; \nonumber \\
& \forall q\in \{\emptyset, R, B, RB\}^T & \nonumber
\end{align}
which can be solved numerically. However, the number of states grows exponentially in $T$ which makes the characterization of strategies and computation of the optimal strategy complicated especially for large $T$. For small $T$, we might be able to find the optimal solution explicitly. To illustrate the structure of the optimal solution, we present the explicit solutions for $T=0$ and $T=1$ in the following two subsections.
\subsection{The Optimal Mix Strategy for $T=0$}
The case of $T=0$ is trivial since, in this case, for stationary policies, the output sequence is i.i.d. as well, and therefore
\begin{eqnarray*}
\mbox{RHS of } \dref{upper2} &=& \frac{1}{N}\sum_{k=1}^N{H(O_k|I_k,G_k)} \nonumber \\
       &=& H(O_1|I_1,G_1) \nonumber \\
      &=& \lambda_R \lambda_B H(O_1|I_1=RB,G_1=2). \nonumber
\end{eqnarray*}
Therefore, to maximize the anonymity, the Mix must send a random permutation of the received packets, in the case of both read and blue arrival, with equal probability to get $H(O_1|I_1=RB,G_1=2)=1$. Correspondingly, the maximum anonymity is given by
$$
A^{\psi^*}=\frac{\lambda_R\lambda_B}{\lambda_R+\lambda_B}.
$$

In the rest of this section, we consider the more interesting case of $T=1$, where each packet has to be sent out in the current slot or in the next slot.
\subsection{The Optimal Mix Strategy for $T=1$}
As we proved in section \ref{structure}, we only need to consider the class of stationary policies that maximize the RHS of \dref{upper2}. Therefore, for $T=1$, the optimal mixing strategy is the solution to the following average entropy maximization problem
$$
\max \lim _{N \to \infty} \frac{1}{N}\sum_{k=1}^N{E_{q_{k-1}}\left[H(O_k|I_k,q_{k-1},G_k)\right]},
$$
where now $q_{k-1} \in \{\emptyset, R, B, RB\}$. Recall that the random variable $Q_{k-1}$ (=$Q_{k-1}(0)$ here) denotes what has been left in the queue, at the end of time slot $k-1$, for transmission in the time slot $k$, where we have defined the initial condition as $Q_0=\emptyset$, and $q_{k-1}$ is the realization of $Q_{k-1}$. Roughly speaking, the action $\psi_k$ is to randomly select some packets from $I_k$ and $q_{k-1}$, and send the permutation of the selected packets to the output. Let $w$ denote the maximum value of the above average entropy maximization problem, then, by definition, $A^{\psi^*}=\frac{w}{\lambda_R+\lambda_B},$
and the optimal mixing strategy $\psi^*$ is the one that chooses the corresponding optimal policy for the average entropy maximization problem. In order to solve the problem, next we identify the possible actions for different states which will allow us to define the reward function in more detail and provide an explicit solution.

\subsubsection{Set of possible actions and corresponding rewards for different states}\label{sub1}
There is a set of possible actions for each state depending on different arrival types. In the following, we identify the set of actions and their corresponding rewards for each case.
\begin{enumerate}[(a)]

\item Assume ${Q_{k-1}=\emptyset}$, then
\begin{itemize}
\item [(i)] If $I_k=\emptyset$: In this case, obviously, there will be no transmission at the output link, $G_k=0$, and the queue will remain empty as well, i.e.,  $O_k=\emptyset$ and $Q_k=\emptyset$. The corresponding entropy is $H(O_k|I_k=\emptyset,Q_{k-1}=\emptyset, G_k)=0$.
\item [(ii)] If $I_k=R$: Two options are possible; the Mix can queue the arrived packet ($G_k=0$) with probability $\alpha_k$, or send the packet in the current slot ($G_k=1$) with probability $1-\alpha_k$. No matter what the Mix does, the entropy in this slot $H(O_k|I_k=R,Q_{k-1}=\emptyset,G_k)=0$. Correspondingly, the queue is updated as $Q_k=R$, with probability of $\alpha_k$, or $Q_k=\emptyset$, with probability of $1-\alpha_k$.
\item [(iii)] If $I_k=B$: This case is similar to the previous case except that we use $\beta_k$ instead of $\alpha_k$. Therefore, $Q_k=B$, with probability $\beta_k$, or $Q_k=\emptyset$, with probability $1-\beta_k$, and $H(O_k|I_k=B,Q_{k-1}=\emptyset,G_k)=0$.
\item [(iv)] If $I_k=RB$: The Mix has four options; it can queue both packets (with probability $1-s_k$), send both out (with probability $s_k(1-y_k)$), keep only $R$ and send $B$ out (with probability $s_ky_k(1-p_k)$), or keep only $B$ and send $R$ out (with probability $s_ky_kp_k$). Note that the parameters $s_k$, $y_k$, and $p_k$ have been used to characterize the probabilities of different options. Intuitively, $s_k$ is the probability that a transmission at the output link happens at all, $y_k$ is the probability of sending only one packet out given a transmission must happen, and $p_k$ is the probability of sending $R$ out given that only one packet is transmitted at the output. Accordingly,
    \be
    H(O_k|I_k=RB,Q_{k-1}=\emptyset,G_k) \nonumber \\
    =s_k\left( y_k\mathcal{H}(p_k)+1-y_k\right), \nonumber
    \ee
    where $\mathcal{H}$ is the binary entropy function given by
    $$\mathcal{H}(p)=-p\log(p)-(1-p)\log(1-p)$$
    for $0<p<1$.
\end{itemize}

\item Assume $Q_{k-1}=R$, then
\begin{itemize}
\item [(i)] If $I_k=\emptyset$: The Mix has to send the content of the queue to the output, therefore $O_k=R$, and obviously, $H(O_k|I_k=\emptyset,Q_{k-1}=R,G_k)=0$ and $Q_k=\emptyset$.
\item [(ii)] If $I_k=R$: The Mix can queue the recent $R$, with probability $\gamma_k$, and send $Q_{k-1}$ to the output, or can send both $Q_{k-1}$ and the recent arrival to the output, with probability $1-\gamma_k$. Therefore, $Q_k=R$ ($O_k=R$) with probability $\gamma_k$, or $Q_k=\emptyset$ ($O_k=RR$) with probability $1-\gamma_k$. The corresponding entropy will be zero, i.e., $H(O_k|I_k=R,Q_{k-1}=R,G_k)=0$.
\item [(iii)] If $I_k=B$: Again the Mix has two options; it can send a random permutation of $R$ and $B$ to the output, i.e., $Q_k=\emptyset$, with probability $a_k$, or it can queue the $B$ and send only the $R$ out, i.e., $Q_k=B$, with probability $1-a_k$. The entropy is $H(O_k|I_k=B,Q_{k-1}=R,G_k)=a_k$.
\item [(iv)] If $I_k=RB$: The Mix has three options; it can queue both arrivals, i.e., $Q_k=RB$,  with probability $1-t_k$, keep only the red arrival in the queue, i.e., $Q_k=R$, with probability $t_k(1-d_k)$, or keep only the blue arrival in the queue, i.e., $Q_k=B$, with probability $t_kd_k$. Correspondingly, in this case,
\be
 P(O_k=o_k|I_k=RB,Q_{k-1}=R,G_k=2) \nonumber \\
 =
    \left\{
      \begin{array}{ll}
       d_k &;o_k= RR  \\
       (1-d_k)/2 &; o_k= RB  \\
       (1-d_k)/2 &; o_k= BR.
      \end{array}
    \right.
\ee
and
$$H(O_k|I_k=RB,Q_{k-1}=R,G_k)=t_k\left( \mathcal{H}(d_k)+1-d_k\right).$$
\end{itemize}

\item Assume ${Q_{k-1}=B}$, then this case is similar to the previous case and the details are omitted for brevity.
\begin{itemize}
\item [(i)] If $I_k=\emptyset$: Obviously, $H(O_k|I_k=\emptyset,Q_{k-1}=B,G_k)=0$, and $Q_k=\emptyset$.
\item [(ii)] If $I_k=B$: $H(O_k|I_k=B,Q_{k-1}=B,G_k)=0$. Options are $Q_k=B$, with probability $\delta_k$, or $Q_k=\emptyset$, with probability $1-\delta_k$.
\item [(iii)] If $I_k=R$: $H(O_k|I_k=R,Q_{k-1}=B,G_k)=b_k$. Options are $Q_k=R$, with probability $1-b_k$, or $Q_k=\emptyset$, with probability $b_k$.
\item [(iv)] If $I_k=RB$: The Mix can keep both arrivals in the queue, i.e., $Q_k=RB$, with probability $1-z_k$, keep only the red arrival in the queue, i.e., $Q_k=R$, with probability $z_k r_k$, or keep only the blue arrival in the queue, i.e., $Q_k=B$, with probability $z_k(1-r_k)$. The entropy is $$H(O_k|I_k=RB,Q_{k-1}=B,G_k)=z_k\left( \mathcal{H}(r_k)+1-r_k\right).$$
\end{itemize}

\item Assume ${Q_{k-1}=RB}$, then the Mix has to send the contents of the queue to the output, i.e., $O_k=RB$ or $BR$ with equal probabilities, and queue all the recent arrivals, i.e., $Q_k=I_k$. The entropy is simply $H(O_k|I_k,Q_{k-1}=RB,G_k)=1$.
\end{enumerate}

Next, we calculate the reward for each state. Recall that the reward function is
$$
C(x_k, \psi_k)=H(O_k|I_k,q_{k-1},G_k)=E_{i_k}\left[H(O_k|i_k,q_{k-1},G_k)\right],
$$
where $i_k$ denotes a realization of $I_k$. Therefore, averaging over $4$ possible arrivals in each state, the reward function and queue updates, for each state are the following.
\begin{enumerate}[(a)]

\item $Q_{k-1}=\emptyset$:

The reward function is given by
\begin{equation*}
C(\emptyset, \psi_k)=\lambda_R \lambda_B s_k\left( y_k\mathcal{H}(p_k)+1-y_k\right),
\end{equation*}
and the queue is updated as
\begin{eqnarray*}
&& P(Q_k=q|Q_{k-1}=\emptyset, \psi_k)=\\
&& \left\{\begin{array}{ll}
\lambda_R(1-\lambda_B)\alpha_k+\lambda_R \lambda_B s_ky_k(1-p_k)&; q=R \\
\lambda_B(1-\lambda_R)\beta_k+\lambda_R \lambda_B s_k y_k p_k &;q=B\\
\lambda_R \lambda_B (1-s_k) & ;q=RB\\
--- &; q=\emptyset
\end{array}
\right.
\end{eqnarray*}
where we used the notation $``---"$ for the probability of having an empty queue, since we will not need the explicit expression for this probability, although, it can be, obviously, derived from the other three probabilities. Note that $\psi_k$ is specified by $5$ parameters $0\leq \alpha_k$, $\beta_k$, $y_k$, $s_k$, $p_k\leq1$.

\item $Q_{k-1}=R$:

The reward function is given by
\begin{equation*}
C(R, \psi_k)=\lambda_B(1-\lambda_R)a_k+\lambda_R\lambda_B t_k\left( \mathcal{H}(d_k)+1-d_k\right),
\end{equation*}
and the queue is updated as
\begin{eqnarray*}
&& P(Q_k=q|Q_{k-1}=R, \psi_k)=\\
&& \left\{\begin{array}{ll}
\lambda_R(1-\lambda_B)\gamma_k+\lambda_R \lambda_B t_k(1-d_k)&;q=R \\
\lambda_B(1-\lambda_R)(1-a_k)+\lambda_R \lambda_B t_k d_k&;q=B \\
\lambda_R \lambda_B (1-t_k)&;q=RB\\
---&;q=\emptyset
\end{array}
\right.
\end{eqnarray*}
Note that, in this state, $\psi_k$ is specified by $4$ parameters $0\leq a_k$, $t_k$, $\gamma_k$, $d_k \leq1$.
\item $Q_{k-1}=B$:

The reward function is given by
\begin{equation*}
C(B, \psi_k)=\lambda_R(1-\lambda_B)b_k+\lambda_R\lambda_B z_k\left( \mathcal{H}(r_k)+1-r_k\right),
\end{equation*}
and the queue is updated as
\begin{eqnarray*}
&& P(Q_k=q|Q_{k-1}=B, \psi_k)= \\
&& \left\{\begin{array}{ll}
\lambda_R(1-\lambda_B)(1-b_k)+\lambda_R \lambda_B r_kz_k&;q=R\\
\lambda_B(1-\lambda_R)\delta_k+\lambda_R \lambda_B z_k (1-r_k)&;q=B\\
\lambda_R \lambda_B (1-z_k)&;q=RB\\
---&;q=\emptyset
\end{array}
\right.
\end{eqnarray*}
Note that here $\psi_k$ is specified by $4$ parameters $0\leq b_k$, $z_k$, $\delta_k$, $r_k \leq1$.
\item $Q_{k-1}=RB$:

The reward function is given by
\begin{equation*}
C(RB, \psi_k)=1,
\end{equation*}
and the queue is updated as
\begin{eqnarray*}
&& P(Q_k=q|Q_{k-1}=RB, \psi_k)=\\
&& \left\{\begin{array}{ll}
\lambda_R(1-\lambda_B)&;q=R\\
\lambda_B(1-\lambda_R)&;q=B\\
\lambda_R \lambda_B &;q=RB \\
---&;q=\emptyset
\end{array}
\right.
\end{eqnarray*}
Note that, here, there is no degrees of freedom for $\psi_k$ (The Mix has to send out $Q_{k-1}$).
\end{enumerate}

\subsubsection{The Optimal Stationary Mix strategy}\label{sub2}
Having formally defined the reward function and the dynamics of the system in subsection \ref{sub1}, we use Lemma \ref{mdp} to solve the average reward maximization problem. It turns out that the optimal strategy is specified by only three parameters $p$, $r$, and $d$, and all the other parameters must be one. The following proposition states one of our main results.
\begin{proposition}\label{main}
For the double input-single output Mix, and $T=1$, the optimal Mix strategy is the following. At each time $k$, given $Q_{k-1}$ and $I_k$, if
\begin{enumerate}
\item $Q_{k-1}=\emptyset$
\begin{itemize}
\item $I_k= \emptyset$, $R$, $B$: $Q_k=I_k$, $O_k=\emptyset$.
\item $I_k=RB$: send $R$ out with probability $p^*$ or $B$ with probability $1-p^*$, $Q_k=I_k \backslash O_k$.
\end{itemize}
\item $Q_{k-1}=R$
\begin{itemize}
\item $I_k=\emptyset, R$: $Q_k=I_k$, $O_k=Q_{k-1}$.
\item $I_k=B$: transmit a random permutation of $R$ and $B$, $Q_k=\emptyset$.
\item $I_k=RB$: transmit $RR$ with probability $d^*$ ($Q_k=B$), or transmit a random permutation of $R$ and $B$ with probability $1-d^*$ ($Q_k=R$).
\end{itemize}
\item $Q_{k-1}=B$
\begin{itemize}
\item $I_k=\emptyset, B$: $Q_k=I_k$, $O_k=Q_{k-1}$.
\item $I_k=R$: transmit a random permutation of $R$ and $B$, $Q_k=\emptyset$.
\item $I_k=RB$: transmit $BB$ with probability $r^*$ ($Q_k=R$), or transmit a random permutation of $R$ and $B$ with probability $1-r^*$ ($Q_k=B$).
\end{itemize}
\end{enumerate}
where probabilities $p^*$, $d^*$, and $r^*$ depend on arrival rates $\lambda_R$ and $\lambda_B$.
\end{proposition}
In the special case $\lambda_R=\lambda_B$, $p^*=\frac{1}{2}$, $d^*=\frac{1}{3}$, and $r^*=\frac{1}{3}$.
\begin{proof}[Proof of Proposition \ref{main}]
Recall the optimality equation \dref{dp}:
$$
w+\phi(q)=\max_u \left\{C(q,u)+E[\phi(Q_1)|Q_0=q, \psi_1=u]\right\}.
$$
Since $\phi$ is unique up to an additive constant, without loss of generality, assume $\phi(\emptyset)=0$. Then, for $q=\emptyset$, the optimality equation can be written as
\begin{eqnarray*}
w & = & \max_{s,p,y,\alpha,\beta} \left\{
\lambda_R \lambda_B s\left( y\mathcal{H}(p)+1-y\right) \right.\nonumber \\
& & + [\lambda_R(1-\lambda_B)\alpha+\lambda_R \lambda_B s y(1-p)]\phi(R) \nonumber \\
& & + [\lambda_B(1-\lambda_R)\beta+\lambda_R \lambda_B s y p]\phi(B)\nonumber \\
& & + \left. [\lambda_R \lambda_B (1-s)]\phi(RB)
\right\}.
\end{eqnarray*}
Obviously, $\alpha=1$ and $\beta=1$ maximize the right hand side if $\phi(R)$ and $\phi(B)$ are nonnegative. We will later see that $\phi(R)$ and $\phi(B)$ are indeed nonnegative. Therefore, the right hand side of the optimality equation can be written as
\begin{eqnarray*}
\lambda_R\lambda_Bs\left[y\left(\mathcal{H}(p)-1+(1-p)\phi(R)+p\phi(B)\right)+1-\phi(RB)\right]\\
+\lambda_R(1-\lambda_B)\phi(R)+\lambda_B(1-\lambda_R)\phi(B)+\lambda_R\lambda_B\phi(RB).
\end{eqnarray*}
First, consider the term $\mathcal{H}(p)-1+(1-p)\phi(R)+p\phi(B)$. This term is maximized by choosing
\begin{equation}\label{p*}
p^*=\frac{1}{1+2^{\phi(R)-\phi(B)}}.
\end{equation}
We will later show that
\begin{equation}\label{y*cond}
\mathcal{H}(p^*)-1+(1-p^*)\phi(R)+p^*\phi(B) \geq 0,
\end{equation}
and therefore $y^*=1$. Furthermore, for $y^*=1$, we will see that the term inside the brackets is always nonnegative, i.e.,
\begin{equation}\label{wcond}
\mathcal{H}(p^*)+(1-p^*)\phi(R)+p^*\phi(B)-\phi(RB) \geq 0,
\end{equation}
and therefore $s^*=1$. Finally, $w$ is given by
\begin{eqnarray}\label{weq}
w&=&\lambda_R\lambda_B\mathcal{H}(p^*)+\lambda_R(1-\lambda_Bp^*)\phi(R)\nonumber \\
&+&\lambda_B(1-\lambda_R(1-p^*))\phi(B).
\end{eqnarray}
Next, consider the optimality equation for $q=R$. It can be written as
\begin{eqnarray*}
w+\phi(R)&=& \max_{\gamma, d, t, a}\{\lambda_B(1-\lambda_R)a+\lambda_R\lambda_B t\left( \mathcal{H}(d)+1-d\right) \nonumber \\
& &+[\lambda_R(1-\lambda_B)\gamma+\lambda_R\lambda_B(1-d)]\phi(R) \nonumber \\
& & +[\lambda_B(1-\lambda_R)(1-a)+\lambda_R\lambda_Btd]\phi(B) \nonumber \\
& &+\lambda_R\lambda_R(1-t)\phi(RB)\}.
\end{eqnarray*}
Similar to the argument for $q=\emptyset$, $\gamma^*=1$, if $\phi(R)>0$, and $a^*=1$ if $\phi(B)<1$. Furthermore, taking the derivative respect to $d$, setting it to zero, and solving it for $d^*$ yields
\begin{equation}\label{d*}
d^*=\frac{1}{1+2^{1+\phi(R)-\phi(B)}}.
\end{equation}
Finally, $t^*=1$ if
\begin{equation}\label{d*cond}
\mathcal{H}(d^*)+1-d^*+(1-d^*)\phi(R)+d^*\phi(B)-\phi(RB) \geq 0,
\end{equation}
and the optimality condition is simplified to
\begin{eqnarray}\label{d*eq}
w+\phi(R)&=& \lambda_B(1-\lambda_R)+\lambda_R\lambda_B \left( \mathcal{H}(d^*)+1-d^*\right) \nonumber \\
& & +[\lambda_R(1-\lambda_B)+\lambda_R\lambda_B(1-d^*)]\phi(R)\nonumber \\
& & +\lambda_R\lambda_Bd^*\phi(B).
\end{eqnarray}
Next, consider the optimality equation for $q=B$
\begin{eqnarray*}
w+\phi(B)&=& \max_{\delta, r, z, b}\{\lambda_R(1-\lambda_B)b+\lambda_R\lambda_B z\left( \mathcal{H}(r)+1-r\right) \nonumber \\
& &+[\lambda_B(1-\lambda_R)\delta+\lambda_R\lambda_Bz(1-r)]\phi(B) \nonumber \\
& & +[\lambda_R(1-\lambda_B)(1-b)+\lambda_R\lambda_Bzr]\phi(R) \nonumber \\
& &+\lambda_R\lambda_R(1-z)\phi(RB)\}.
\end{eqnarray*}
In parallel with the argument for $q=R$, $\delta^*=1$ if $\phi(B) \geq 0$, and $b^*=1$ if $\phi(R) \leq 1$. Moreover, $z^*=1$ if
\begin{equation}\label{r*cond}
\mathcal{H}(r^*)+1-r^*+(1-r^*)\phi(B)+r^*\phi(R)-\phi(RB) \geq 0,
\end{equation}
where
\begin{equation}\label{r*}
r^*=\frac{1}{1+2^{1+\phi(B)-\phi(R)}}.
\end{equation}
The optimality condition is simplified to
\begin{eqnarray}\label{r*eq}
w+\phi(B)&=& \lambda_R(1-\lambda_B)+\lambda_R\lambda_B \left( \mathcal{H}(r^*)+1-r^*\right) \nonumber \\
& &+[\lambda_B(1-\lambda_R)+\lambda_R\lambda_B(1-r^*)]\phi(B) \nonumber \\
& &+\lambda_R\lambda_Br^*\phi(R).
\end{eqnarray}

Finally, the optimality equation for $q=RB$ is given by
\begin{eqnarray}\label{RBeq}
w+\phi(RB)&=& 1+\lambda_R(1-\lambda_B)\phi(R) \nonumber \\
& & +\lambda_B(1-\lambda_R)\phi(B)+\lambda_R\lambda_B\phi(RB)
\end{eqnarray}
Therefore, we need to solve equations (\ref{weq}), (\ref{d*eq}), and (\ref{r*eq}) to find $w$, $\phi(R)$, and $\phi(B)$. Then, (\ref{RBeq}) can be used to find $\phi(RB)$. Eventually, what remains to be shown is that $0 \leq \phi(R), \phi(B) \leq 1$, and, in addition, $\phi(R)$, $\phi(B)$, and $\phi(RB)$ satisfy inequalities (\ref{y*cond}), (\ref{wcond}), (\ref{d*cond}), and (\ref{r*cond}).

First, consider the special case of $\lambda_R=\lambda_B=\lambda$. By symmetry, $\phi(R)=\phi(B)$ which yields $p^*=1/2$ and $d^*=r^*=1/3$. Then, by solving equations (\ref{weq}) and (\ref{d*eq}), we have,
\begin{equation*}\label{phiR}
\phi(R)=\phi(B)=\frac{\lambda^2(\log 3-2)+\lambda}{-\lambda^2+\lambda+1},
\end{equation*}
and
\begin{equation*}\label{wsimp}
w=\frac{\lambda^2}{-\lambda^2+\lambda+1}\left[-\lambda^2(\log 3-1)+2(\log 3 -2)\lambda+3\right].
\end{equation*}
Then, the anonymity is $A^{\psi^*}=w/2\lambda$, and it is easy to check that the solutions satisfy all the inequalities.
Figures \ref{wesp}\ and \ref{phiResp} show the anonymity $A^{\phi^*}$ and $\phi(R)$ as functions of $\lambda$.
\begin{figure}[!t]
\centering
\subfigure[Anonymity]{\label{wesp}
\includegraphics [width = 3 in]{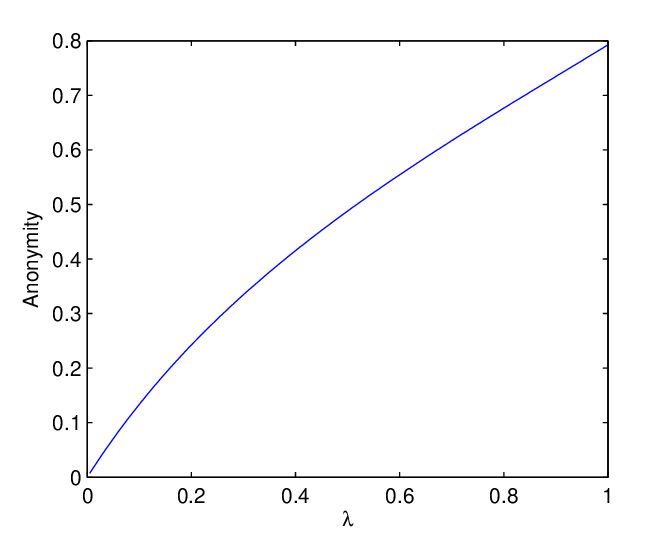}
}
\subfigure[$\phi(R)$]{\label{phiResp}
\includegraphics [width = 3 in]{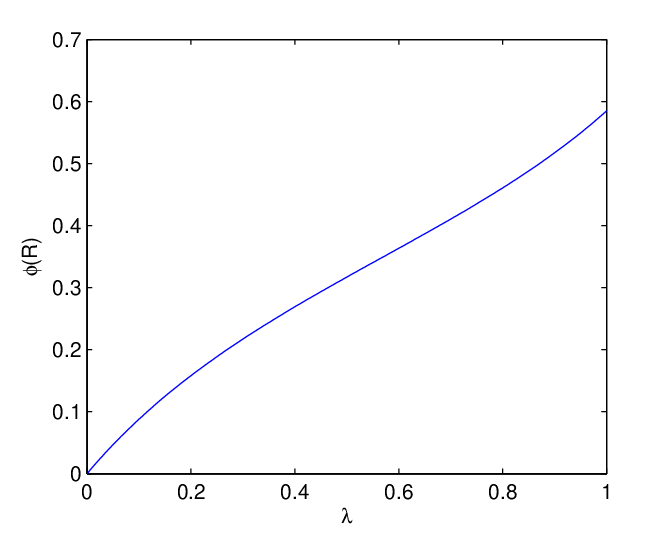}
}
\label{wphiResp}
\caption{Anonymity and $\phi(R)$ for the case of $\lambda_R=\lambda_B=\lambda$.}
\end{figure}

Next, consider the general case with, probably, unequal arrival rates. We prove that the solutions indeed exist and they satisfy the required conditions. Using (\ref{weq}) to replace $w$ in (\ref{d*eq}) and (\ref{r*eq}) yields
\begin{eqnarray}
\phi(R)&=&\lambda_B[1-\phi(B)(1-\lambda_R)] +\lambda_R\lambda_Bg(\xi),
\label{sys1}
\end{eqnarray}
\begin{eqnarray}
\phi(B)&=&\lambda_R[1-\phi(R)(1-\lambda_B)] +\lambda_R\lambda_Bf(\xi),
\label{sys2}
\end{eqnarray}
where,
$$
g(\xi)=(d^*-p^*)(-\xi)+ \mathcal{H}(d^*) - \mathcal{H}(p^*)-d^*,
$$
$$
f(\xi)=(r^*+p^*)\xi+ \mathcal{H}(r^*) - \mathcal{H}(p^*)-r^*-\xi ,
$$
and,
$$ \xi=\phi(R)-\phi(B).$$

Therefore, the optimal probabilities can be expressed as functions of $\xi$ by
\begin{eqnarray*}
p^*&=& \frac{1}{1+2^\xi}, \\
d^*&=&\frac{1}{1+2^{1+\xi}}, \\
r^*&=&\frac{1}{1+2^{1-\xi}}.
\end{eqnarray*}
\begin{lemma}\label{gf}
The function $g(\xi)$ is an increasing function of $\xi$ and $f(\xi)$ is a decreasing function of $\xi$ (see Appendix for the proof).
\end{lemma}
For any pair $(\phi(R), \phi(B))$ chosen from $[0,1] \times [0,1]$, we have $-1 \leq \xi \leq 1$, and therefore, by Lemma \ref{gf}, functions $f$ and $g$ can be bounded from below and above by
$$
g(-1) \leq g(\xi) \leq g(1),
$$
and
$$
f(1) \leq f(\xi) \leq f(-1).
$$
But it is easy to check that
$$
g(1)=f(-1)=\log (5/3)-1 , \ g(-1)=f(1)=1-\log{3},
$$
and therefore,
$$-1 < f(\xi),g(\xi) < 0.$$
Consequently, the right-hand sides of (\ref{sys1}) and (\ref{sys2}) form a continuous mapping from $[0,1] \times [0,1]$ to $[0,1] \times [0,1]$, and therefore, by the Brouwer fixed point theorem (\cite{brouwer}, p. 72), the system of nonlinear equations, (\ref{sys1}), (\ref{sys2}), has a solution $(\phi(R), \phi(B)) \in [0,1] \times [0,1]$.
%

Next, we show that the solutions indeed satisfy the inequalities.
First, we prove that (\ref{y*cond}) holds. Define
\begin{eqnarray*}
\psi_1 (\xi)&= & \mathcal{H}(p^*)+(1-p^*)\phi(R)+p^*\phi(B) \nonumber \\
& =& \mathcal{H}(p^*) -p^* \xi +\phi(R).
\end{eqnarray*}
First, consider the case that $-1 \leq \xi \leq 0$, then
\begin{eqnarray*}
\frac{d}{d\xi}(\mathcal{H}(p^*)-p^*\xi)&=&{p^*}^\prime \log \frac{1-p^*}{p^*}-p^*-{p^*}^\prime \xi \nonumber \\
& =&-p^* \leq 0.
\end{eqnarray*}
Hence,
$$
\psi_1(\xi) \geq \psi_1(0) =1+ \phi(R) \geq 1.
$$
For the case that $0 \leq \xi \leq 1$, rewrite $\psi_1(\xi)$ as the following
$$
\psi_1(\xi)=\mathcal{H}(p^*)+(1-p^*) \xi +\phi(B).
$$
Then,
\begin{eqnarray*}
\frac{d}{d\xi}(\mathcal{H}(p^*)+(1-p^*)\xi)=1-p^* \geq 0,
\end{eqnarray*}
and hence,
$$
\psi_1(\xi) \geq \psi_1(0) =1+ \phi(B) \geq 1.
$$
Therefore, for $-1 \leq \xi \leq 1$, $\psi_1(\xi) \geq 1$, and (\ref{y*cond}) holds.

Note that from (\ref{RBeq}), we have
\begin{equation}\label{RBeq2}
\phi(RB) =\frac{1-\lambda_R\lambda_B\psi_1(\xi)}{1-\lambda_R\lambda_B},
\end{equation}
and since (\ref{y*cond}) holds, we have
\begin{equation*}
\phi(RB) \leq 1,
\end{equation*}
and consequently (\ref{wcond}) will be satisfied as well.

To show (\ref{d*cond}), note that $\phi(R)+1-\phi(RB) \geq 0$, and therefore, it suffices to prove that
\begin{eqnarray*}
\psi_2(\xi)&=&\mathcal{H}(d^*)-d^*-d^*\phi(R)+d^*\phi(B)\nonumber \\
& =& \mathcal{H}(d^*)-d^*\xi -d^*
\end{eqnarray*}
is nonnegative. But $\psi_2(\xi)$ is a decreasing function since
\begin{eqnarray*}
\frac{d}{d\xi}\psi_2&=&{d^*}^\prime \log {\frac{1-d}{d}} -{d^*}^\prime \xi -d^*-{d^*}^\prime \nonumber \\
& =& {d^*}^\prime (1+\xi) -{d^*}^\prime \xi -d^*-{d^*}^\prime \nonumber \\
& =& -d^* \leq 0
\end{eqnarray*}
So $\psi_2(\xi) \geq \psi_2(1) = \mathcal{H}(1/5)-2/5=\log 5 -2 \geq 0$, and consequently (\ref{d*cond}) follows.
(\ref{r*cond}) is also proved by a similar argument. Define a function $\psi_3(\xi)$ as
\begin{eqnarray*}
\psi_3(\xi)&=&\mathcal{H}(r^*)-r^*-r^*\phi(B)+r^*\phi(R)\nonumber \\
& =& \mathcal{H}(r^*)+r^*\xi -r^*.
\end{eqnarray*}
Then, $\psi_3(\xi)$ is an increasing function since
\begin{eqnarray*}
\frac{d}{d\xi}\psi_3=r^* \geq 0.
\end{eqnarray*}
Thus,
\begin{eqnarray*}
\psi_3(\xi) & \geq & \psi_3(-1)\\
& = & \mathcal{H}(1/5)-2/5\\
&=& \log 5 -2 \geq 0,
\end{eqnarray*}
and therefore (\ref{r*cond}) follows. This concludes the proof of Proposition \ref{main}.
\end{proof}
\subsubsection{Numerical results}
Equations (\ref{weq}), (\ref{d*eq}), and (\ref{r*eq}) form a system of nonlinear equations which can be solved numerically, for different values of $\lambda_R$ and $\lambda_B$, by using the following algorithm.
\begin{algorithm}
\caption{}
\begin{algorithmic}[1]
\STATE $p^*_{0} \leftarrow1/2$, $d^*_{0} \leftarrow 1/3$, $r^*_{0} \leftarrow 1/3$
\STATE $i \leftarrow 0$
\REPEAT
\STATE $i \leftarrow (i+1)$
\STATE $w, \phi(R), \phi(B)$ $\Leftarrow$ solve (\ref{weq}), (\ref{d*eq}), and (\ref{r*eq})
\STATE $p^*_{i}, d^*_{i}, r^*_{i}$ $\Leftarrow$ calculate (\ref{p*}), (\ref{d*}), and (\ref{r*})
\UNTIL{$|p^*_{i}-p^*_{i-1}| \leq \epsilon$ and $|d^*_{i}-d^*_{i-1}| \leq \epsilon$ and $|r^*_{i}-r^*_{i-1}| \leq \epsilon$}
\end{algorithmic}
\end{algorithm}

\begin{figure}[!t]
\centering
\includegraphics [width = 3.5 in]{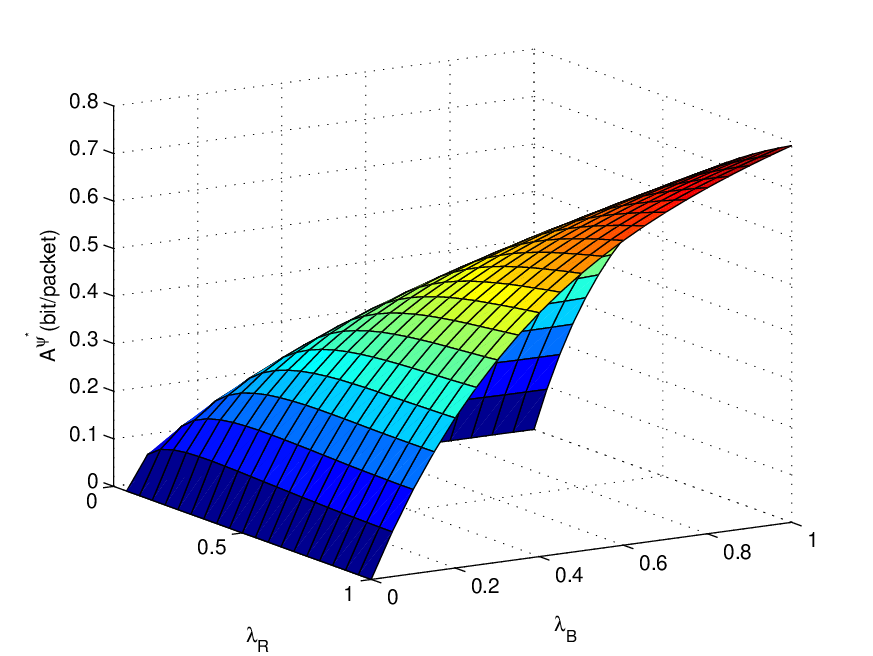}
\caption{Anonymity for different values of $\lambda_R$ and $\lambda_B$.}
\label{optimal-anon}
\end{figure}

\begin{figure}[!t]
\centering
\subfigure[$p^*$]{
\includegraphics [width = 3 in]{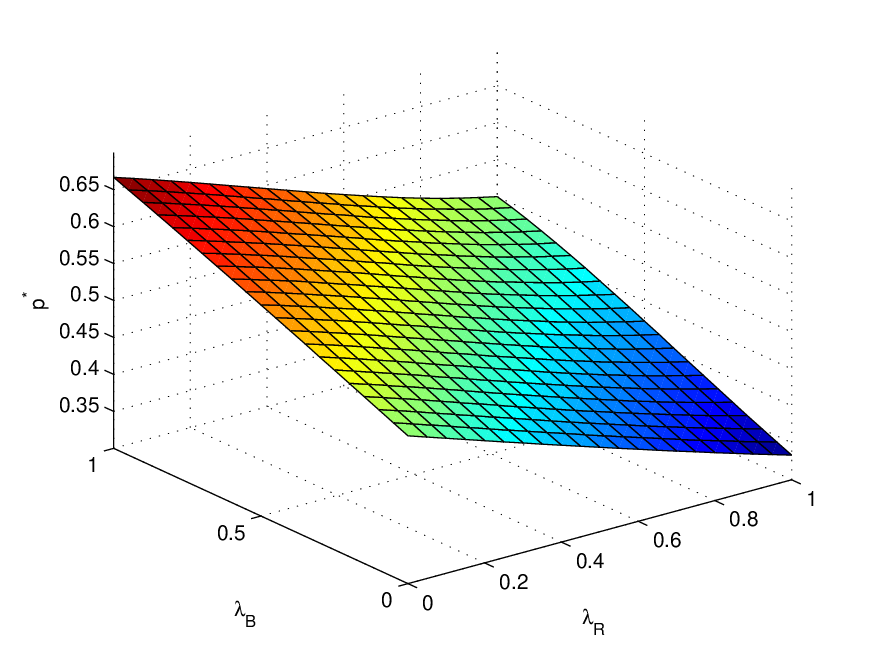}
}
\hspace{0.05 in}
\subfigure[$d^*$]{
\includegraphics [width = 3 in]{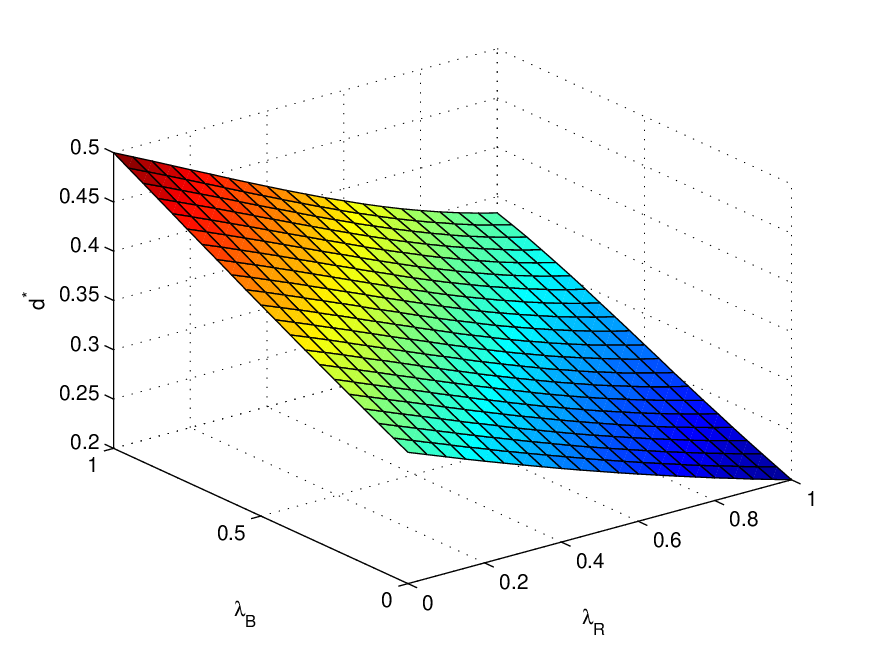}
}

\subfigure[$r^*$]{
\includegraphics [width = 3 in]{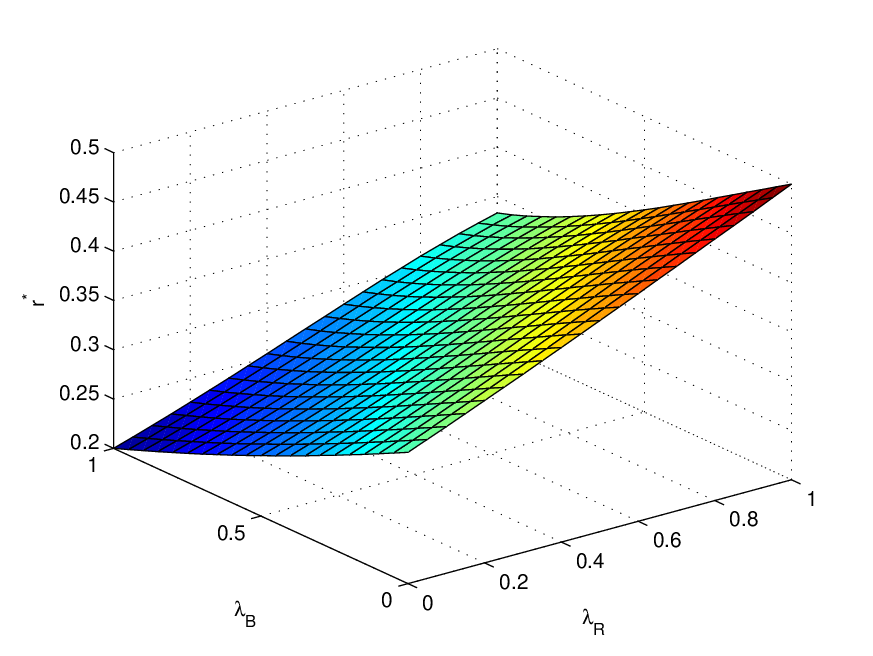}
}
\caption{Probabilities $p^*$, $d^*$, and $r^*$ for different arrival rates $\lambda_R$ and $\lambda_B$.}
\label{probs}
\end{figure}

Note that in the step 5 of the algorithm, we solve a linear system of equations ($p^*$, $d^*$, and $r^*$ are replaced with their numerical values).
Figure \ref{optimal-anon} shows the maximum anonymity, found by running the algorithm, for different arrival rates.
The probabilities $p^*$, $d^*$, and $r^*$ of the optimal mixing strategy have been evaluated in Figure \ref{probs} for different arrival rates $\lambda_R$ and $\lambda_B$.

\begin{remark}
The stationary policy does not exist for $\lambda_R=\lambda_B=1$ since as $\lambda_R \to 1$ and $\lambda_B \to 1$, $\phi(RB) \to -\infty$ (see (\ref{RBeq2})). This makes sense since, in this case, if we start with initial condition $Q_0=\emptyset$ and use the strategy specified in Proposition \ref {main}, we get an anonymity of $A^{\psi^*}=\log{(3)}/2$; whereas if the initial condition is $Q_0=RB$, the only possible strategy will be to transmit the contents of the queue, and queue the arrived $RB$ in each time slot. This yields an anonymity of $1/2$ bit/packet. Therefore, the optimal strategy depends on the initial condition for $\lambda_R=\lambda_B=1$.
\end{remark}
\section{Double Input-Double Output Mix}\label{double-double}
Figure \ref{mixer2} shows the double input-double output Mix. The capacity of each link is $1$ packet/time slot. Compared to the Mix with one output link, i.e., Figure \ref{mixer}, the flows of outgoing packets are separate.

At this point, we would like to clarify the main difference between Sections \ref{double-single} and \ref{double-double} of the paper. The focus of both sections is on flow-level anonymity. However, in the double input-single output Mix, section \ref{double-single}, every packet can be analyzed to see if it belongs to a particular flow. In the double input-double output Mix, even if one packet is identified as belonging
to a flow, then it compromises the entire flow at that node. Hence, in this case, the eavesdropper does not need to detect the sender for each outgoing packet; instead, it aims to find the corresponding source of each flow, by observing a sequence of outgoing packets of sufficiently long duration. Let $d_R \in \{1,2\}$ denote the destination of the red source. Formally, we define the anonymity $A^\psi$ of a Mix strategy $\psi$ for the double input-double output Mix to be
\be
A^\psi=\lim_{N \to \infty} H\big(d_R|I_R^{(N)}, I_B^{(N)}, G_1^{(N)}, G_2^{(N)}\big),
\ee
where similar to Section \ref{double-single}, $I_R^{(N)}$ and $I_B^{(N)}$ are the sequences of red and blue arrivals of length $N$, and $G_i^{(N)}=(G_i(1), G_i(2),...,G_i(N))$, $i=1,2$, where $G_i(t) \in \{0,1\}$ indicates whether there is a packet at the output $i$ in time slot $t$.
Without loss of generality, assume that $\lambda_R > \lambda_B$ (the singular case of $\lambda_R=\lambda_B$ will be discussed later). Then, by calculating the long-run average rates of outgoing flows, the eavesdropper can identify the corresponding source-destination pairs. Therefore, it is not possible to get any anonymity without dropping some packets from the red flow. Hence, the maximum achievable throughput for each flow cannot be more than $\min \{\lambda_R, \lambda_B \}(=\lambda_B)$, and, at least, the packets of the flow with higher rate, which is the red flow here, must be dropped at an average rate of $\lambda_R-\lambda_B$.

We now present our model for a Mix with two queues for red and blue arrivals. Let $A_R[0,t]$ and $A_B[0,t]$ be the number of arrivals for the red and blue arrivals in $[0,t]$. Also let $D_1[0,t]$ and $D_2[0,t]$ denote the number of departures from the output links $1$ and $2$ by the end of time slot $t$. Then, to assure any nonzero anonymity, i.e, we need
\begin{equation}\label{inequality}
D_i[0,t] \leq \min\{A_R[0,t], A_B[0,t]\}; \ \forall t \geq 1; \mbox{ for } i=1,2
\end{equation}
This is clear, because, for example, if there exists a time $t_1$ such that $A_R[0,t_1] \geq D_1[0,t_1] > A_B[0,t_1]$, then obviously the red source is connected to the output $1$ and the anonymity is zero.
\subsection{Mix under a strict delay constraint $T$}
 Suppose that each arrival has to be transmitted within $T$ time slots. In this case, in addition to red packets, blue packets have to be dropped as well. This is because it might happen that there is a blue arrival but no red packets for a time duration of $T$, in which case transmitting the blue packet will immediately reveals the corresponding destination of the blue source. Hence, the throughput of each flow will be less $\lambda_B$. Recall the queue model of the Mix described earlier and consider the following strategy.

  \textit{Mix strategy under strict delay $T$:} For each queue, transmit the \textit{head-of-the-line} (HOL) packet  along with the HOL packet of the other queue simultaneously at the corresponding output links. If the HOL packet has been in the queue for more than $T$ time slots and there is no arrivals in the other queue, drop the HOL packet.
  \begin{proposition}
  The above strategy is optimal in the sense that it yields perfect anonymity with maximum possible throughput.
  \end{proposition}
  Perfect anonymity, means that $A^{\psi}=1$, i.e., by observing the output sequence, the eavesdropper cannot obtain any information and each outgoing flow is equally likely to belong to one of sources.
  \begin{proof}
  Noting that any strategy with non-zero anonymity must satisfy \dref{inequality}, it is easy to observe that packets that are dropped under our strategy will be dropped under any strategy that satisfy \dref{inequality}. Hence, our strategy has the maximum throughput.
Clearly our strategy has also perfect anonymity because $G_1(t)=G_2(t)$ at all times $t$. Also note that in our strategy, packets will be immediately transmitted once a different color packet appears in the other queue. This is the earliest time that a packet can be transmitted under \dref{inequality}. Hence, the average delay of those packets transmitted successfully is also minimized under our strategy.
\end{proof}
Next, consider the case that there is no strict delay constraint. In this case, one does not know how to measure the average delay because any Mix with non-zero anonymity has to drop some of the packets and the delay of the dropped packets is infinity. So, instead, we use the average queue size as the QoS metric.
\subsection{Mix with an average queue length constraint}
In this case, instead of a per-packet latency constraint, we consider the average queue size as the QoS metric.
Similar to the previous case, we consider strategies that achieve both maximum throughput and perfect anonymity. Among all such strategies, we will find an optimal strategy that minimizes the mean queue length.

First note that, to get the smallest queue size, we would like (\ref{inequality}) to hold with equality, i.e.,
\begin{equation}\label{equality}
D_1[0,t] = D_2[0,t] = \min\{A_R[0,t], A_B[0,t]\}; \ \forall t \geq 1
\end{equation}
Then, it is clear that red and blue packets must be transmitted simultaneously on output links, i.e., red packets are only transmitted when there is a blue packet in the second queue, and similarly, the blue packets are served when there is a red packet in the first queue.

Also note that dividing both sides of (\ref{equality}) by $t$ and taking the limit as $t \to \infty$ shows that the maximum throughput should be $\min\{\lambda_R, \lambda_B\}$. Therefore, the optimal strategy must drop the Red packets at an average rate $\lambda_R- \lambda_B$, in a way that minimizes the mean queue length, while retaining equality (\ref{equality}).

 Next, consider the problem of minimizing the mean queue length. This problem can be posed as an infinite-state Markov decision problem with unbounded cost. It follows from checking standard conditions, e.g., \cite{lippman}, \cite{linn}, that a stationary optimal policy exists for our problem, however, the average-cost optimality equation \dref{dp} may not hold. Therefore, we follow a different approach.

\begin{figure}[!t]
\centering
\includegraphics [width = 3 in]{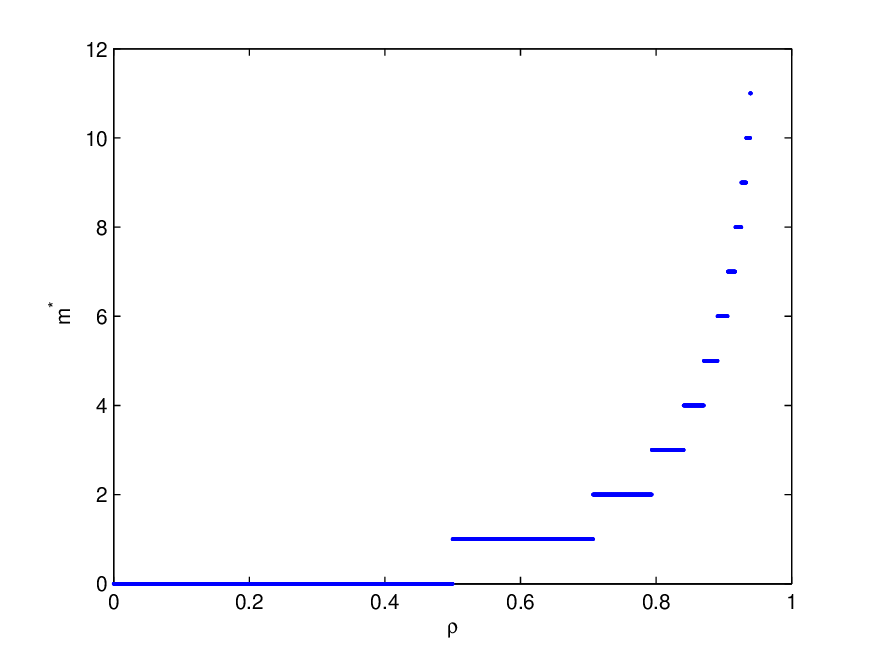}
\caption{The optimal threshold to minimize the average queue length.}
\label{threshold}
\end{figure}

Recall that when a red packet and a blue packet are both available, to minimize queue length, it is best to transmit them immediately. Therefore, when one of the queues (blue or red) hits zero, from that point onwards, only one of the queues can be non-empty. Thus in steady-state, we can assume that one queue can be non-empty. As a result, we have the Markov decision process described next.
Let $(i,j)$ represent the state of the system where there are $i$ packets in the red queue and $j$ packets in the blue queue. The transition probabilities are given by
\begin{eqnarray*}
P[(0,y)|(0,y)]&= & \lambda_R\lambda_B+(1-\lambda_R)(1-\lambda_B) \\
P[(0,y-1)|(0,y)]& =& \lambda_R(1-\lambda_B)\\
P[(0,y+1)|(0,y)]& =& \lambda_B(1-\lambda_R),
\end{eqnarray*}
and
\begin{eqnarray*}
P[(x,0)|(x,0)]&= &\lambda_R\lambda_B+(1-\lambda_R)(1-\lambda_B) \\
 & + & \lambda_R(1-\lambda_B)\delta_x\\
P[(x-1,0)|(x,0)]&=&\lambda_B(1-\lambda_R)\\
P[(x+1,0)|(x,0)]&=& \lambda_R(1-\lambda_B)(1-\delta_x),
\end{eqnarray*}
where $\delta_x$ denotes probability of dropping the red packet in state $(x,0)$, if there is a red arrival but no blue arrival. Note that stability of the system (finiteness of mean queue length) implies that the red packets must be dropped at an average rate of $\lambda_R -\lambda_B$.
So our problem is to determine $\delta_x$ for each $x$ to minimize the mean queue length. We will show that the optimal policy is a threshold policy, which is defined below.
\begin{definition}
A threshold policy, with threshold $m$, is a policy that has the following properties: $\delta_x=0$ for all $0 \leq x \leq m-1$, and $\delta_{m}=1$, where $m$ is a nonnegative integer number.
\end{definition}
The following proposition presents the main result regarding the optimal strategy.
\begin{proposition}\label{th2}
For the double input-double output Mix, the threshold policy is optimal, in the sense that it minimizes the average queue size among all maximum throughput policies with perfect anonymity. Moreover, the threshold is given by
\begin{equation}
m^*=\left\{\begin{array}{ll}
    \lceil {- \frac{1}{\log{\rho}}} \rceil -1 &; \frac{1}{2} < \rho < 1 \\
      0 &; 0 \leq \rho \leq \frac{1}{2},
      \end{array}
    \right.
\end{equation}
where $
\rho=\frac{\lambda_B(1-\lambda_R)}{\lambda_R(1-\lambda_B)}.
$
\end{proposition}
In other words, no buffer is needed for $\lambda_R \geq \frac{2\lambda_B}{1+\lambda_B}$, but, as rates get closer, for $\lambda_B < \lambda_R<\frac{2 \lambda_B}{1+\lambda_B}$, a buffer of size $m^*$ for the red flow is needed.
The optimal threshold $m^*$ is depicted in Figure \ref{threshold}. Note that the singular case of $\lambda_R=\lambda_B=\lambda$ ($\rho=1$) is not stable. By allowing a small drop rate of $\epsilon \lambda$ for each flow, where $0 <\epsilon \ll 1$, one buffer for each flow can be considered, and the thresholds and the average queue size can be expressed as functions of $\epsilon$.
\begin{proof}[Proof of Proposition \ref{th2}]
The steady state distribution for the Markov chain representing the double input-double output Mix is given by
\begin{eqnarray*}
\pi_{0,y}&=&\pi_{0,0}\rho^y, \ y=1,2, \cdots \\
\pi_{x,0}&=&\pi_{0,0}\rho^{-x} \prod_{i=0}^{x-1} (1-\delta_i), \ x=1,2, \cdots
\end{eqnarray*}
where
$$
\pi_{0,0}=\left(\frac{1}{1-\rho}+\sum_{x=1}^\infty \rho ^{-x} \prod_{i=0}^{x-1} (1-\delta_i)\right)^{-1},
$$
and
$$
\rho=\frac{\lambda_B(1-\lambda_R)}{\lambda_R(1-\lambda_B)}.
$$
Recall that, by assumption, $\lambda_R >\lambda_B$, and therefore $ 0 \leq \rho <1$. The average queue length is
\begin{eqnarray*}
\bar{L}&=&\sum_{y=0}^\infty y \pi_{0,y}+\sum_{x=1}^{\infty}x\pi_{x,0}\\
 & =& \pi_{0,0} \left[\frac{\rho}{(1-\rho)^2}+\sum_{x=1}^\infty x\rho ^{-x} \prod_{i=0}^{x-1} (1-\delta_i) \right].
\end{eqnarray*}
Note that for any nonnegative integer $j$, and for fixed values of $\delta_i$s, $i \neq j$, $\bar{L}$ is a linear fractional function of $\delta_j$. More formally,
$$
\bar{L}(\delta_j)=\frac{A_j+(1-\delta_j)B_j}{A_j^\prime+(1-\delta_j)B_j^\prime},
$$
where
\begin{eqnarray*}
A_j^\prime&=&\frac{1}{1-\rho}+\sum_{x=1}^j \rho ^{-x} \prod_{i=0}^{x-1} (1-\delta_i),\\
A_j&=&\frac{\rho}{(1-\rho)^2}+\sum_{x=1}^j x \rho ^{-x} \prod_{i=0}^{x-1} (1-\delta_i), \\
B_j^\prime &=&\frac{\prod_{i=0}^{j-1}(1-\delta_i)}{\rho^{j+1}}\left[1+\sum_{x=1}^\infty \rho ^{-x} \prod_{i=j+1}^{x+j} (1-\delta_i) \right],\\
\end{eqnarray*}
and
$$
B_j=\frac{\prod_{i=0}^{j-1}(1-\delta_i)}{\rho^{j+1}}\left[j+1+\sum_{x=1}^\infty (j+x+1)\rho ^{-x} \prod_{i=j+1}^{x+j} (1-\delta_i) \right].
$$
Therefore, $\partial {\bar{L}}/\partial {\delta_j}$ is either positive or negative, independent of $\delta_j$, and consequently, the optimal $\delta_j$ to minimize $\bar{L}$ is either $0$ or $1$, i.e., $\delta^*_j \in \{0,1\}$ for all $j$. But, all of the $\delta_j$s cannot be zero, otherwise the system will not be stable ($\bar{L}=\infty$). Define $m$ to be the smallest $j$ such that $\delta^*_j=1$. Then $\delta_x=0$ for all $0 \leq x \leq m-1$, and $\delta_{m}=1$ which yields a threshold policy with threshold $m$. Therefore the threshold policy is the optimal policy.

Next, we find the optimal threshold $m^*$. The stationary distribution of a threshold policy with threshold $m$ is given by
\begin{eqnarray*}
\pi_{0,y}&=&\pi_{0,0}\rho^y, \ y=1,2, \cdots \\
\pi_{x,0}&=&\pi_{0,0}(1/ \rho)^x, \ x=1,2, \cdots, m
\end{eqnarray*}
where $\pi_{0,0}=(1-\rho)\rho^{m}$. Therefore, $\pi_{m,0}=1-\rho$, and the average packet-drop rate, $P_{drop}$, is given by
$$
P_{drop}=\pi_{m,0}\lambda_R(1-\lambda_B)=\lambda_R-\lambda_B
$$
which is independent of the threshold $m$. The average queue length is given by
\begin{eqnarray}\label{qlength}
\bar{L}(m)&=& \sum_{y=1}^\infty y \pi_{0,y}+\sum_{x=0}^{m} x \pi_{x,0} \nonumber \\
&=&\left(2 \rho^{m+1}+m(1-\rho)-\rho\right)/(1-\rho).
\end{eqnarray}
Note that $\bar{L}(m)$, as a continuous function of $m$, is strictly convex over $m \in [0,\infty)$ for any fixed $0 \leq \rho <1$; therefore, it has a unique minimizer $m^*$ which is either zero or the solution of $\frac{\partial \bar{L}}{\partial m}=0$. Since we seek the smallest integer-valued $m^*$, the convexity implies that $m^*$ is zero if
$$
\bar{L}(0) \leq \bar{L}(1),
$$
or it's a positive integer $m^*$ satisfying
$$
\bar{L}(m^*) < \bar{L}(m^*-1),
$$
and
$$
\bar{L}(m^*) \leq \bar{L}(m^*+1).
$$
Then by using (\ref{qlength}), it follows that $m^*=0$ if $\rho \leq \frac{1}{2}$, and for $\rho > \frac{1}{2}$, it satisfies
$$
2 \rho^{m^*} > 1,
$$
and
$$
2 \rho^{m^*+1} \leq 1,
$$
which yields
$$
m^*= \lceil {- \frac{1}{\log{\rho}}} \rceil -1 .
$$
This concludes the proof.
\end{proof}
\begin{remark}
As far as the average queue size is concerned, it does not matter which packet is dropped when $\delta_x=1$. However, in order to get a better delay performance for those packets that are not dropped, it is better to accept the new arrival and drop the head-of-the line packet.
\end{remark}
\section{Conclusions}
The definition of anonymity and the optimal mixing strategy for a router in an anonymous network depend on its functionality. In the case of a double input-single output Mix, an eavesdropper knows the next hop of every packet but the router attempts to hide the identity of the packet at the output link so as to make it harder for the eavesdropper to follow the path of a flow further downstream. On the other hand, when there are two inputs, two outputs and only two flows, even revealing the identity of one packet at the output compromises that portion of both flow's route. For the first case, the optimal mixing strategy was found to achieve the maximum anonymity under a per-packet latency constraint. For the second case, the maximum throughput strategies with perfect anonymity were found for a per-packet latency constraint and for minimum average queue size.
Our results in this paper represent a first attempt at theoretically characterizing optimal mixing strategies in two fundamental cases. Further research is needed to find optimal mixing strategies under more general constraints or for the multiple input-multiple output Mix.
\appendices
\section{}
\begin{proof}[Proof of Lemma \ref{gf}]
Taking the derivative of $g$ respect to $\xi$ yields
\begin{eqnarray*}
g^\prime(\xi) & = & (p^*-d^*)+({p^*}^\prime-{d^*}^\prime)\xi+{d^*}^\prime \frac{d}{d d^*}\mathcal{H}(d^*) \\
 & & -{p^*}^\prime \frac{d}{d p^*}\mathcal{H}(p^*)-{d^*}^\prime,
\end{eqnarray*}
but
$$
\frac{d}{d d^*}\mathcal{H}(d^*)=\log{\frac{1-d^*}{d^*}}=1+\xi
$$
and
$$
\frac{d}{d p^*}\mathcal{H}(p^*)=\log{\frac{1-p^*}{p^*}}=\xi,
$$
therefore
$$
g^\prime(\xi)=(p^*-d^*)
$$
which is always nonnegative for all values of $\xi$.
Similarly for $f(\xi)$, we have
\begin{eqnarray*}
f^\prime(\xi) & = & (r^*+p^*)+({p^*}^\prime+{r^*}^\prime)\xi+{r^*}^\prime \frac{d}{d r^*}\mathcal{H}(r^*)\\
& - & {p^*}^\prime \frac{d}{d p^*}\mathcal{H}(p^*)-{r^*}^\prime-1,
\end{eqnarray*}
but
$$
\frac{d}{d r^*}\mathcal{H}(r^*)=\log{\frac{1-r^*}{r^*}}=1-\xi,
$$
and, as we saw,
$$\frac{d}{d p^*}\mathcal{H}(p^*)=\xi,$$
therefore
\begin{eqnarray}
f^\prime(\xi)&=&r^*+p^*-1 \nonumber \\
& =& \frac{1}{1+2^{1-\xi}} +\frac{1}{1+2^\xi}-1 \nonumber \\
& = & \frac{2^\xi}{2^\xi+2}+\frac{1}{1+2^\xi}-1 \nonumber \\
& \leq & \frac{2^\xi}{1+2^\xi}+\frac{1}{1+2^\xi}-1 \nonumber \\
& =& 0.
\end{eqnarray}
Hence, $f(\xi)$ is a decreasing function.
\end{proof}

\end{document}